\documentclass[a4paper, 12pt]{article}
\usepackage{geometry, graphicx, setspace, floatflt} 
\title{Cold-shaping of thin glass foils as novel method for mirrors processing. From the basic concepts to mass production of mirrors.}

\begin{document}
\date{}
\maketitle
\pagenumbering{arabic} 

\begin{flushleft}
Reviewed on September 12th, 2012.\\
Accepted for publication on October 23rd, 2012.\\
To be published on Optical Engineering, Vol 52, No 5, May 2013, Special edition on \textit{Optical Materials}.\\
\end{flushleft}

\vspace{1cm}

\begin{center}
\textbf{Rodolfo Canestrari and Giovanni Pareschi} \\
\textit{INAF - Osservatorio Astronomico di Brera\\
via Emilio Bianchi 46 - 23807 Merate (LC) ITALY\\
phone: +39 0395971044 ; fax: +39 0395971101\\
email: rodolfo.canestrari@brera.inaf.it\\ 
email: giovanni.pareschi@brera.inaf.it}\\ 
\end{center}

\begin{center}
\textbf{Giancarlo Parodi and Francesco Martelli}\\
\textit{BCV progetti s.r.l.\\
via Sant'Orsola 1 - 20123 Milano (MI) ITALY\\
phone: +39 0286452002 ; fax: +39 028900103\\
email: giancarlo.parodi@bcv.it\\
email: martelf@bcv.it}\\
\end{center}

\begin{center}
\textbf{Nadia Missaglia and Robert Banham}\\
\textit{Media Lario Technologies s.p.a.\\
localit\`a Pascolo - 23842 Bosisio Parini (CO) ITALY\\
phone: +39 031867111 ; fax: +39 031876595\\
email: Nadia.Missaglia@media-lario.com\\
email: robert.banham@media-lario.com %\\
}\\ 
\end{center}

\newpage
%\doublespacing

\section*{Abstract}
We present a method for the production of segmented optics. It is a novel processing developed at INAF-Osservatorio Astronomico di Brera (INAF-OAB) employing commercial of-the-shelf materials. It is based on the shaping of thin glass foils by means of forced bending, this occurring at room temperature (cold-shaping). The glass is then assembled into a sandwich structure for retaining the imposed shape. The principal mechanical features of the mirrors are the very low weight, rigidity and environmental robustness. The cost and production time also turns to be very competitive.\\
In this paper we sum up the results achieved during the r\&d performed in the past years. We have investigated the theoretical limits of the structural components by means of parametric finite elements analyses; we also discuss the effects caused by the most common structural loads.
Finally, the process implementation, the more significant validation tests and the mass production at the industry is described.\\

\noindent
\textbf{Keywords:} segmented optics, glass slumping, glass, technology, optical fabrication, mirrors, astronomy

\section{Introduction}
Segmented optics are becoming more and more used in Astronomy because of their capability of coupling low-cost and low-weight together when compared with the monolithic mirror solution. Many different telescopes are already implementing these kind of optics, both ground- and space-based ones. They are either sub-millimeter radio antennas like ALMA or optical telescopes like the Keck, GranTeCan or the forthcoming E-ELT; both the infrared eye of the JWST satellite or the recently launched NuSTAR x-ray telescope mount segmented mirrors.\\
Despite the variety of the scientific topics addressed, the wavelengths of light observed and the technology adopted, there is a commonality between all these (and possibly other) projects. Segmented optics typically require to the manufacturing process the capability to deliver tens or hundreds or even thousands of pieces in a time, cost and requirement frame well defined. Furthermore, each mirror in general is no more a unique piece of high precision optic like in the past but, rather one in a series of identical pieces.\\
There is an additional class of telescopes, newbie in Astronomy, that brings segmented optics into the domain of the extremely low -cost and -weight but with the discount of moderate angular resolution requirement. They are called Imaging Air Cherenkov Telescopes (IACTs). \\
These are instruments for studying astronomical sources that emit very high energy gamma rays. They operate from the ground using Earth's atmosphere as a calorimeter: energetic photons ($\gamma$) or charged particles (hadrons) coming from the most distant parts of the Universe hit the atmosphere and interact with it. They lose energy by producing pairs that run faster than light (in air) and emit a bluish light by the Cherenkov effect. This propagates into the atmosphere, generating a so-called shower, whose light pool covers an area on the ground enclosed by a circle of about 120~m in radius. The electromagnetic showers generated by $\gamma$ are very faint relative to the night-sky background; they are also rare and last for only a few nano-seconds. These light can be detected by very large light collectors (of the order of hundreds of m$^2$) equipped with proper focal plane instrumentation and fast gigahertz electronics. From these images, both the incoming direction and the energy of the primary photons can be recovered. If many telescopes are used and more than one detects the same event, the angular resolution of the reconstructed incoming direction is improved. Such observations make it possible to understand the physics behind the extremely powerful acceleration mechanisms at work in the astronomical sources emitting the primary gamma photons and to gather clues to the origin of the Universe.\\

\begin{figure}[!ht]
%\vspace{5mm}
\centering
\includegraphics[keepaspectratio, scale=1.7]{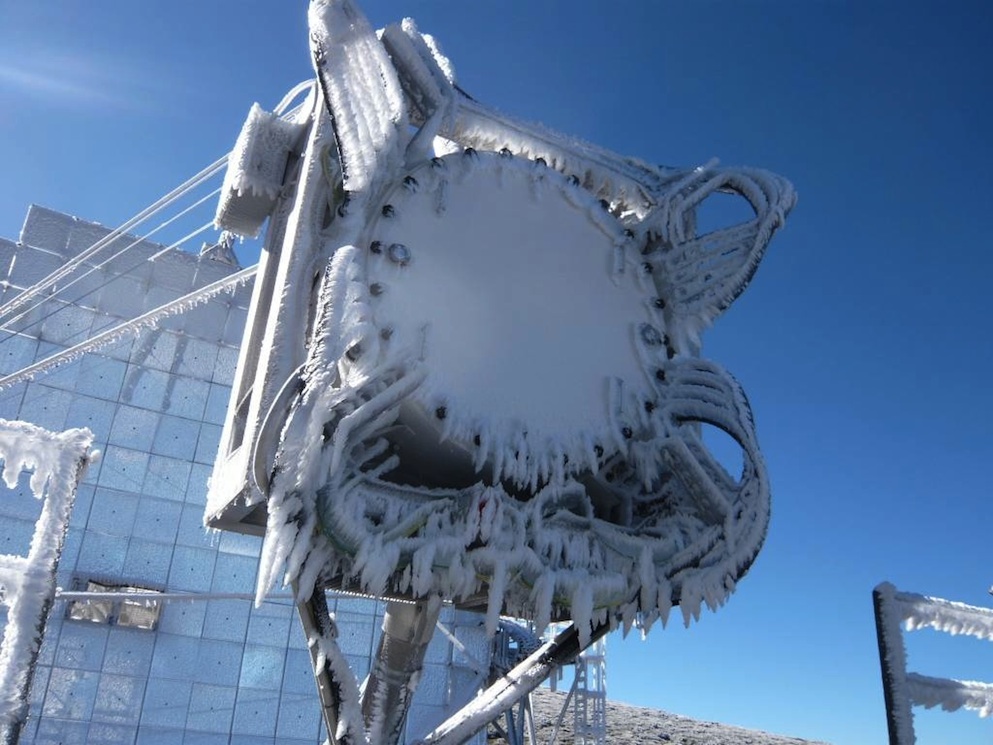}\\
\caption{The MAGIC II telescope during a strong winter season.}
\label{magicghiaccio}
\end{figure}

\noindent
To better exploits this detection technique, IACTs are in general deployed in arrays of few telescopes (2-5 units) working in stereoscopic configuration. They are located on top of mountains, at about 2000-3000 m asl. The sites can hence present aggressive environmental conditions where the telescopes operate: the temperature can ranges from several Celsius degrees below zero to tens above; the humidity can reach 100\%; winds are very frequent and gusts can occur up to 200 km/hr.
Despite these conditions, IACTs are not protected by domes or enclosures. They are continuously exposed to the environment, including the UV solar irradiation, terrain's dust, rain or hail fall, frosting, etc. etc. An example of these is shown in \figurename~\ref{magicghiaccio}. Moreover, the mirrors typically require a cost of simply few thousands of Euro for squared meter and an areal density of 10-30 $kg/m^2$.\\

\section{Concepts of cold-shaping}
\label{concept}

The light collection of IACTs is typically achieved thought one reflection. The collector is tessellated with many \textit{identical} mirror segments; these are sections of a spherical surface. This design is called Davies-Cotton and it has been adapted from early solar concentrators~\cite{dc}. Moreover, as mentioned before, the collecting area of each telescope is very large, also in comparison with the most recent ground-based optical telescopes, and is composed by hundreds of mirror segments. \\
All these aspects underscore the importance and the advantages of the development of proper manufacturing technologies capable of mass production. Processes exploiting the concept of Òreplication of a master shapeÓ can be very attractive to reduce strongly both the costs and the production time while maintaining the quality of the products in terms of repeatability. The cold-shaping (CS) technique here presented has been developed over the past years with the goal to address this problem. Thought this process, a glass foil will copy with a very high degree of fidelity the shape of a mold. The mold can be reused to produce many substrates, one identical to each other, without suffering evident surface deterioration. \\

\begin{figure}[!ht]
%\vspace{5mm}
\centering
\includegraphics[keepaspectratio, scale=1.8]{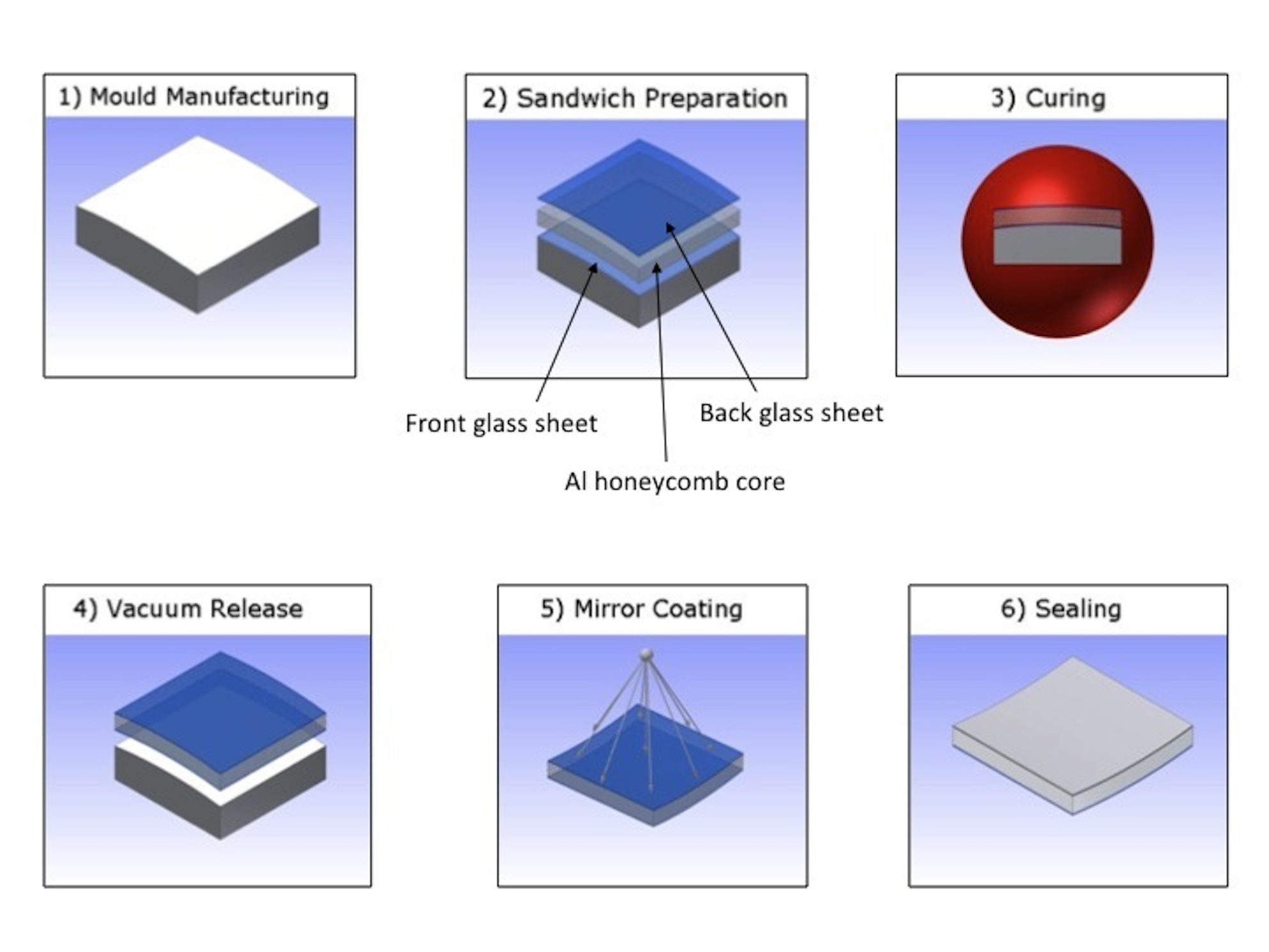}\\
\caption{Conceptual sketches of the main steps for the cold-shaping. Sandwich structural configuration is implemented having a honeycomb core few cm thick and skins made by glass foils.}
\label{cold}
\end{figure}

\noindent
A thin glass sheet, typically 1 or 2 mm thick, is bent by vacuum suction and is made to adhere to a mold. The mold's profile is the negative of that desired on the mirror. This processing is done at room temperature without any heating of the glass. The elasticity of glass, although limited by the tensile strength of the commercial product, is the working principle of this technology. For this reason, clear limitations imposed by this method appear in the surface profiles of the mirrors, particularly the achievable radii of curvature. A detailed analysis is postponed to section~\ref{FEA}. Moreover, since the process occurs at room temperature, the glass permanently retain the tensile stresses resulting from bending. These stresses will also cause the glass to spring back to its original position when the vacuum suction is removed. To overcome this behavior a sandwich-like structural configuration is realized by gluing a reinforcing core and a second glass skin. Aluminum honeycombs, having few cm thickness, are typically used for the core structure. The chosen mechanical structure for the substrate of these mirrors confers stiffness and low areal density.\\
The connection of the parts is achieved through epoxy resin structural adhesive bonding with curing at the temperature required by the glue while maintaining the vacuum suction. After the glue is polymerized, the vacuum suction can be released and the substrate properly coated. Eventually, the mirror is then finished by sealing its edges to prevent damage from water infiltration and ensure its safe handling. \\
\figurename~\ref{cold} shows the main steps as described before.

\section{Structural analyses}
\label{FEA}
Sandwich mirror panels, which consist in two solid face sheets bonded to an inner and lighter core, represent a favorable structural scheme in case it is desired to increase the ratio between the bending stiffness and the panel aerial density.\\
The choice of commercial glass material for the sandwich faceplates is favorable in terms of cost, but it requires a careful evaluation in terms of mechanical strength. As a matter of fact, it is well known that the strength of glass is not an intrinsic property of the material, but on the contrary, it strongly depends on the whole processing method. In fact, strength depends on several parameters such as: distribution of cracks (or surface flaws), entity of stressed surface area (or volume), stress distribution, residual internal stresses from manufacturing process, nature of the loads (static or cyclic), fracture toughness, humidity, temperature. Furthermore, glass materials are susceptible to sub-critical crack growth in monotonic tension due to the influence of moisture. Handling, glass cutting and edge finishing can also affect the surface defects and so the strength. Finally, further strength degradation can occur due to the exposition to the environment.\\
The few above reported remarks point out that a deterministic value for glass strength is not available, the approach to the glass safety is statistical and in principle a proper characterization campaign is necessary. Lacking these information, the main mechanical properties of glass refer to commercial product; specifically for the present analyses we referred to annealed product as in \tablename~\ref{tabglass}.\\
By now the characteristic value of the bending strength, 45 MPa, has been retrieved by engineering judgement as from the most update draft version of the relevant \texttt{prEN 13474-3} regulation~\cite{eurocodes}. Safety checks in mirror design have been performed following the same regulation.\\
We underline the point that in the following sections of the paper we do not aim to present a design for mirrors, instead we discuss some important aspects helpful to drive a design. Nevertheless the methods applied and the finite element models developed give a reliable first order estimate for the stress behavior and can be used for preliminary estimations. \\

\begin{table}[h]
\begin{center}
\begin{tabular}{lll}
\hline
Mass density    		&	& 2.49 g/cm$^3$ \\
Young's modulus    	&	& 73 GPa \\
Poisson's ratio		&	& 0.224 \\
CTE				&	& $9 \times 10^{-6}$ K$^{-1}$ \\
\hline
\end{tabular}
\caption{Mechanical properties of glass as used in the present analyses.}
\label{tabglass}
\end{center}
\end{table}

\subsection{Understanding the bending limits}
Being the glass a brittle material often showing low tensile strength, the stress level in the face sheets become a critical parameter which has to be attentively checked. In the present subsection we focused our attention on the tensile stresses induced by the CS procedure, in order to to assess the limits of the proposed manufacturing procedure. Structural analyses have been carried out by non linear finite element approach, by means of step by step analyses able to follow the stress behavior during the whole bending process as the applied load increases.\\
The purpose of the analyses is to evaluate how different design parameters affect the tensile stress induced by CS. The parameters considered are reported in \tablename~\ref{tabpar}. The analyses performed~\cite{bcvcoldstress} showed that two different configurations, namely \textit{regular} and \textit{corrugated}, of the deformed glass shape can occur during the CS process. The configurations will depend on the tile shape, size and glass thickness as well as on the imposed radius of curvature.\\

\begin{table}[!h]
\begin{center}
\begin{tabular}{lll}
\hline
Tile shape    					&	& square and hexagon \\
Tile area    					&	& 0.6 m$^2$ and 1.2 m$^2$ \\
Glass foils thickness		 		&	& 0.5 - 1.0 - 1.5 - 2.0 mm \\
Radii of curvature of the bending	&	& from 7.5 m to 40 m \\
\hline
\end{tabular}
\caption{Main parameters investigated in finite elements analysis for the CS evaluation.}
\label{tabpar}
\end{center}
\end{table}

\begin{figure}[!ht]
%\vspace{5mm}
\centering
\includegraphics[keepaspectratio, scale=0.9]{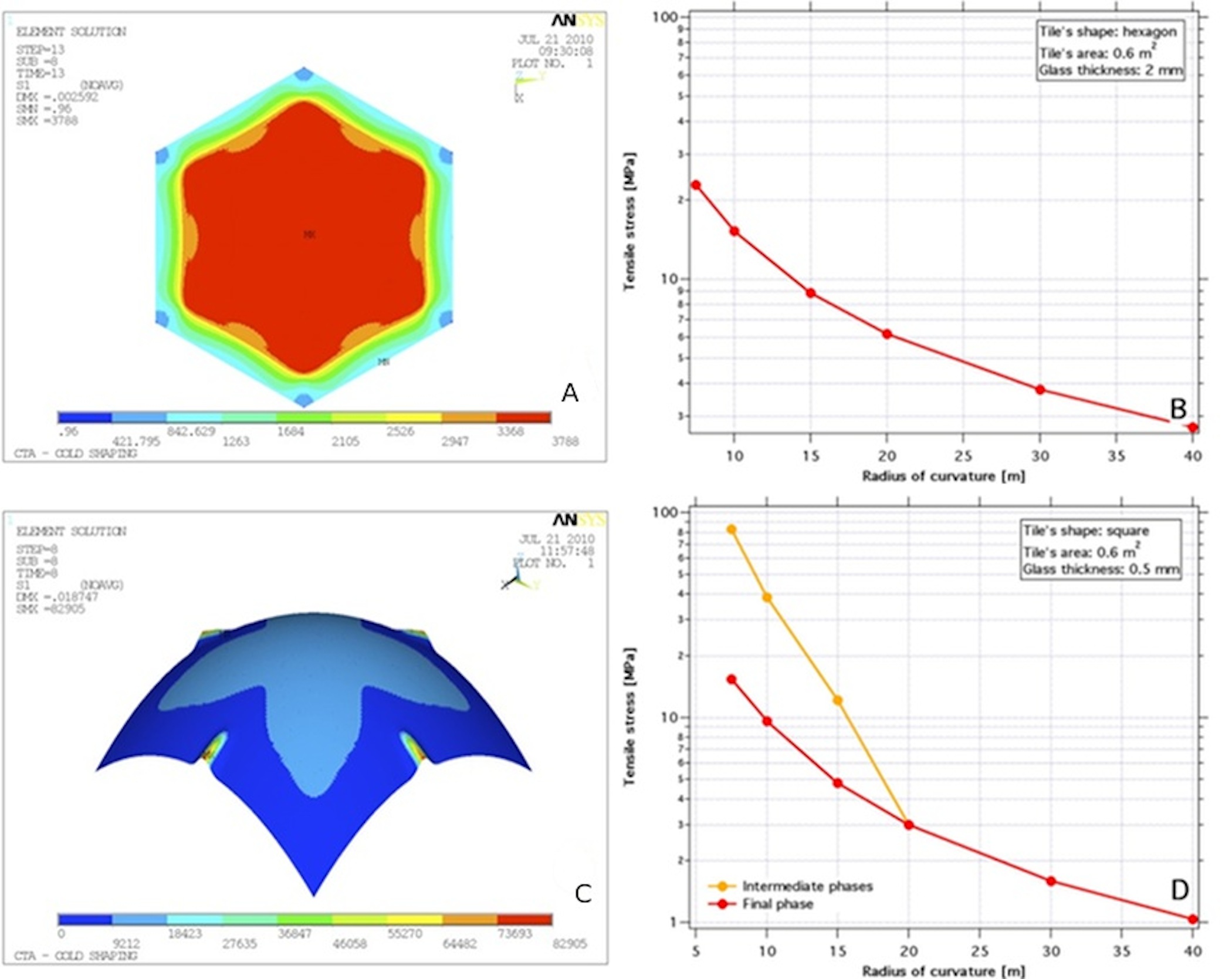}
\caption{Example of CS undergoing the regular configuration: A- isocontours plot of the principal tensile stress; B- principal tensile stress vs. bending radius. Example of CS undergoing the corrugated configuration: C- isocontours plot of the principal tensile stress; D- principal tensile stress vs. bending radius.}
\label{bend-conf}
\end{figure}

\noindent
In the regular configuration the bending of the glass proceeds smoothly and regular from the innermost zones to the outermost ones as the applied pressure increases. No significant folds in the glass are generated during the process. It is the most favorable case in terms of glass stress between the two possible evolutions observed. \figurename~\ref{bend-conf} shows, in the panels A and B, an example of the principal tensile stress developed in the glass during the CS process and the corresponding stress trend as a function of the imposed radius of curvature.\\
On the contrary, in the second case the glass assumes a corrugated configuration where some folds are generated. By increasing the pressure it is in principle possible bringing the glass in contact with the forming mold removing the folds; nevertheless, the intermediate corrugated configurations are very unfavorable in terms of stresses. It is observed that the maximum tensile stress occurs during the intermediate bending phases in correspondence of the folds, while at the end of the process the tensile stresses would be smaller. Any case this is just a theoretical condition, since  the higher stresses recorded could be sufficient to give the failure. \figurename~\ref{bend-conf} C shows the principal tensile stress in the glass during an intermediate phase when the folds are generated. When corrugated configuration arises, a double curve is reported in the plot of \figurename~\ref{bend-conf} D.

\begin{figure}[!ht]
%\vspace{5mm}
\centering
\includegraphics[keepaspectratio, scale=0.95]{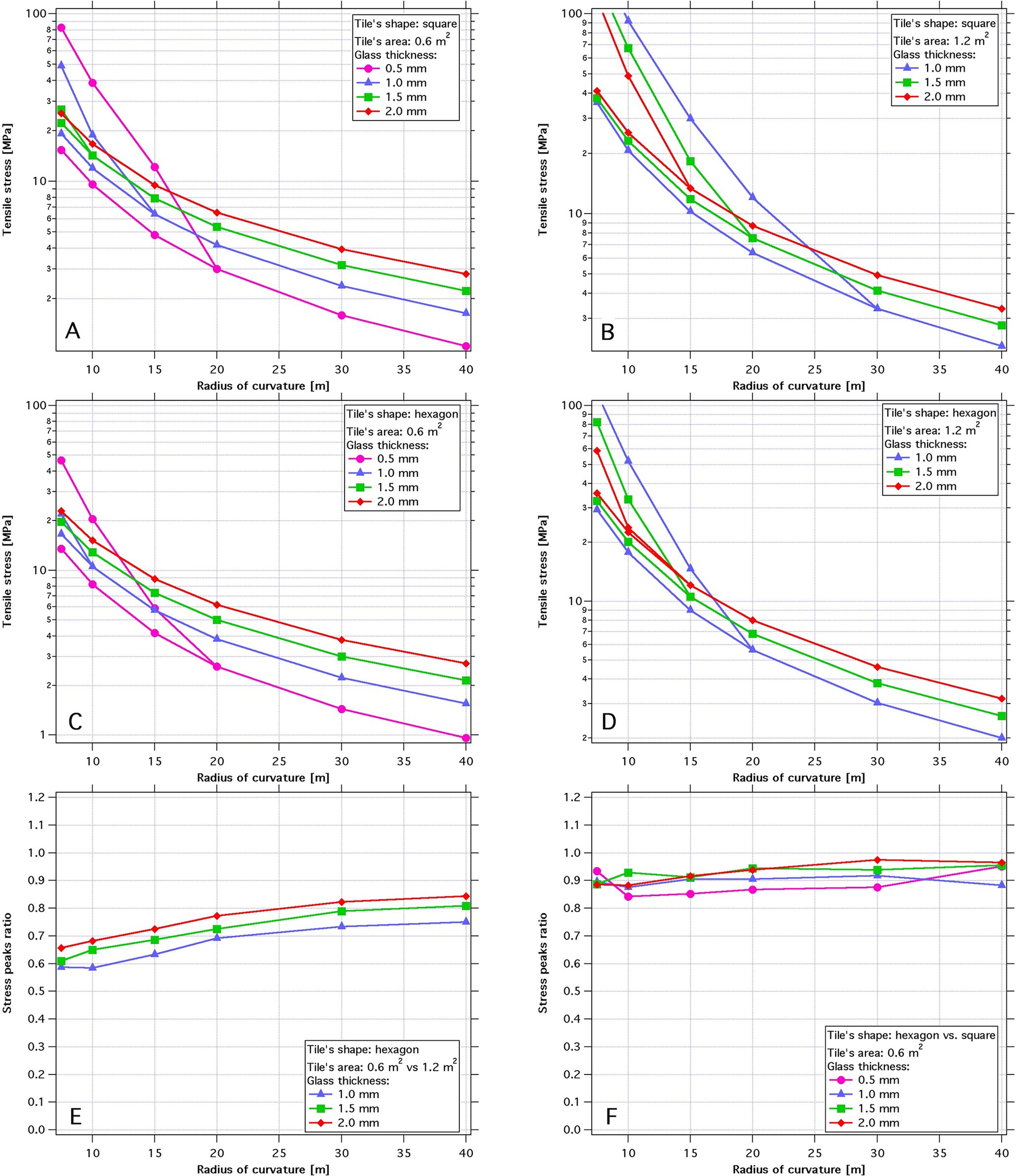}
\caption{Plots showing the behavior of the principal tensile stress vs. bending radius for: A- small square tile; B- large square tile; C- small hexagonal tile and D- large hexagonal tile. Plots showing the relationship between: E- small versus large tiles; F- hexagon versus square tiles.}
\label{stressvsroc}
\end{figure}

\noindent
In \figurename~\ref{stressvsroc} we summarize the tensile stress peaks obtained in all the cases analyzed for the two geometries and sizes considered. From the plots some considerations can be drawn. The formation of folds during the CS process is made easier by: a) reducing the foil thickness; b) increasing the foil size; c) reducing the curvature radius and d) changing the geometry from hexagon to square (under the same panel area). In fact,  by comparing the stress peaks for a given foil thickness and for an imposed radius of curvature, smaller foils seem preferred because stress could be reduced up to about 40\% (see panel E-); similarly, hexagons are slightly better then squares with about 10\% saving in stress (see panel F-). These considerations are worth mentioning in particular for the design of mirrors with small radius of curvature.\\
Finally, we point out that in case of corrugated configurations the upper branch of the curves (intermediate phase) has to be considered just as a first evaluation of the maximum stress, since these analyses only provide results at discrete load steps. It follows that, in principle, there could exist intermediate levels of peak stress that are not revealed by the end case analysis. The same concept is applicable to the bifurcation point of the curves. Nevertheless, since the upper branch of the curve is often related to high and non allowable stress level, the meaning related to this zone of the curve is almost academic.\\

\subsection{Additional effects}
\begin{figure}[!ht]
%\vspace{5mm}
\centering
\includegraphics[keepaspectratio, scale=0.95]{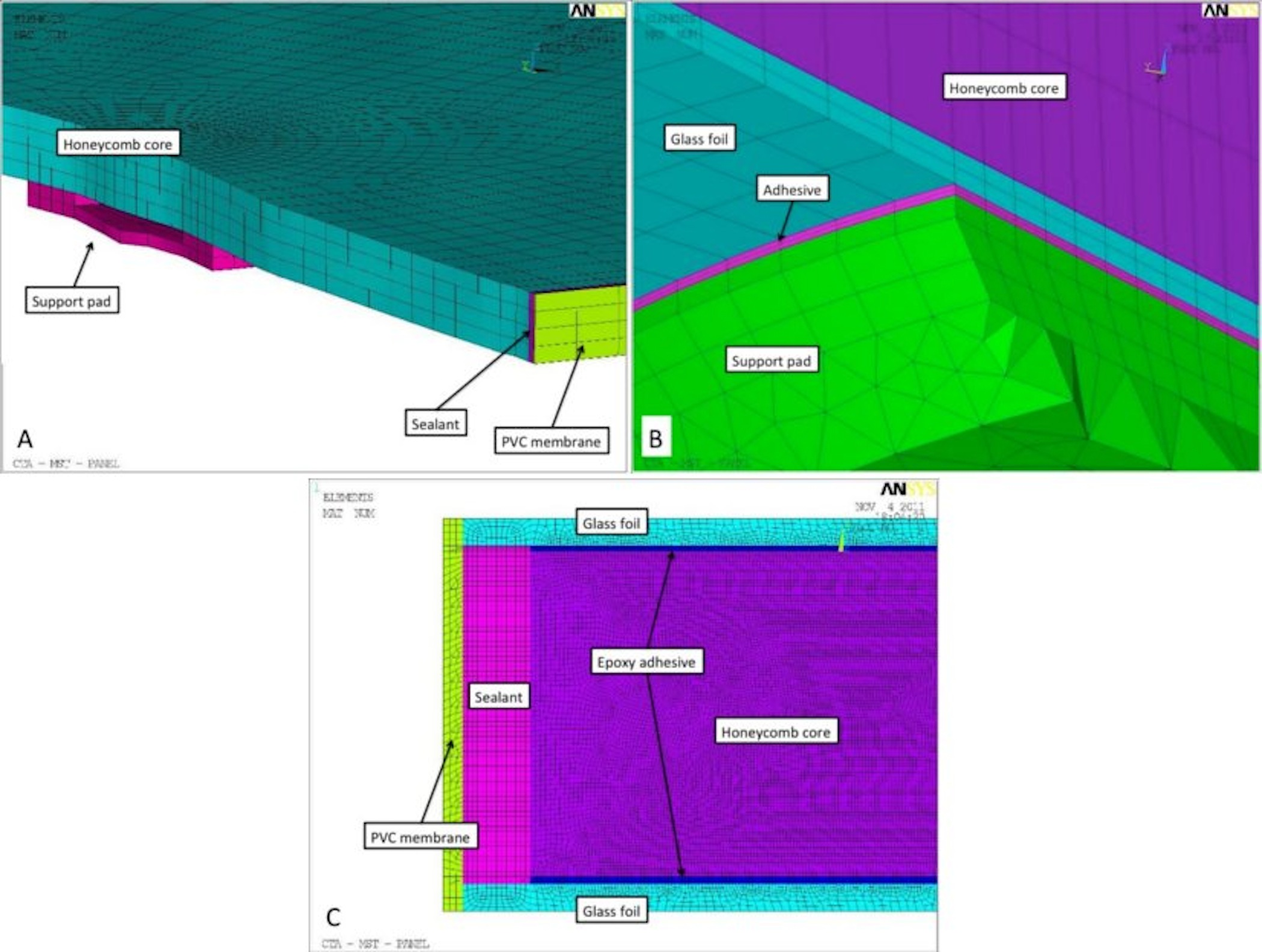}
\caption{Finite Element Models: A- global model for the full panel; B- local model surrounding the pucks locations; C- local model at the mirror edge.}
\label{fem}
\end{figure}

It is evident that the manufacturing process induces stresses in the glass face sheets and possibly in the core material. These stresses are frozen by the bonding between the elements, and so any additional stresses induced by other loads will act on a non stress-free panel. In case the structural behavior of the panel can be considered linear, new stresses induced by any loads will be simply added to the initial stresses related to the CS phase.\\
With reference to the present structural configuration (i.e. sandwich), the influence of the mirror size, core thickness and pucks location have been also investigated with particular attention to:
\begin{itemize}
\item additional stresses due to environmental conditions (i.e. strong winds, large temperature shifts);
\item elastic deformations due to operative conditions (i.e. axial gravity, moderate wind and temperature gradients across panel thickness).
\end{itemize}
Long term effects caused by stress relief in the glass could in principle lead to figure changes. In applications well below the glass transition temperature, as our case, these have been judged negligible since, long-term strains are just a few percent of the instantaneous elastic deformation and so the impact is negligible.\\
Only hexagon tiles have been considered.\\
The deformed shape and the stress state in the panels have been evaluated by means of global finite element models relevant to the whole mirror segment (see panel A- in \figurename~\ref{fem}); more detailed ones have been also implemented to investigate local stress states generated into sensible zones of the mirror as shown by panels B- and C- (i.e. pucks location and mirror edges). A comprehensive selection of the results illustrating the different aspects is reported and discussed in the following.

\begin{figure}[!ht]
\centering
\includegraphics[keepaspectratio, scale=0.9]{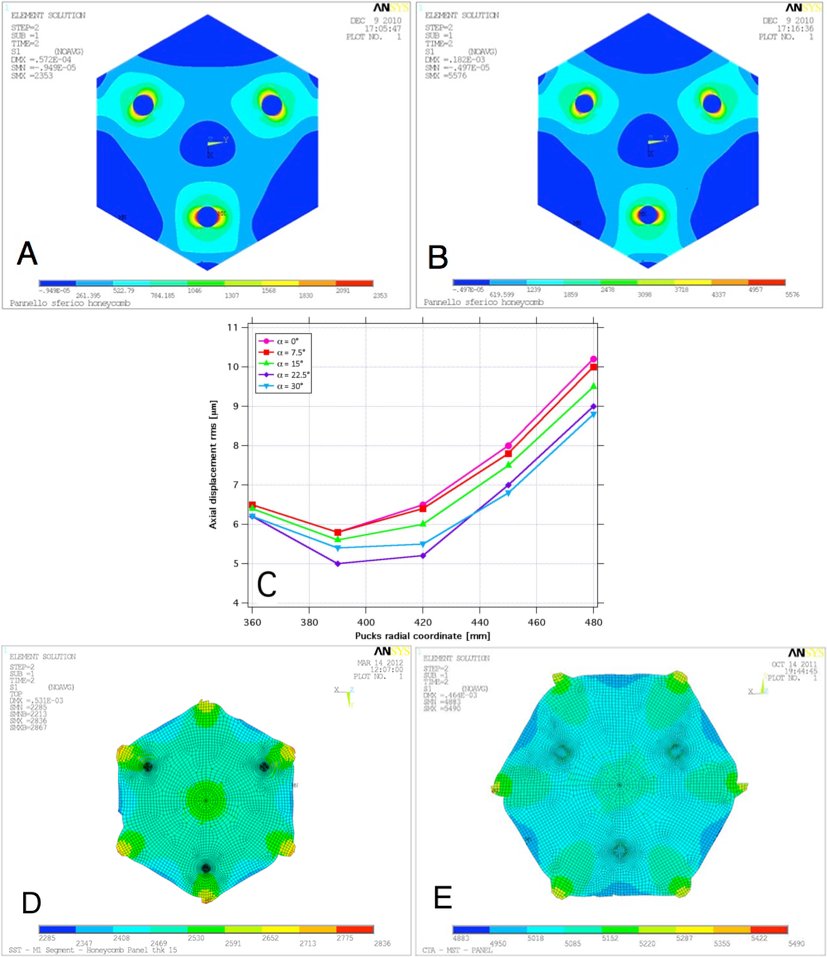}
\caption{Panels A- and B- show respectively examples of principal tensile stress in case of small and large mirrors (core thickness: 30 mm). Panel C- shows the influence of the pucks location for rms errors under normal gravity + moderate winds, $\alpha$ is the azimuthal position of the three pucks. Panels D- and E- show the influence on the stress peaks of a temperature shift of +40$^\circ$C in case of small and large mirrors. Stress isocontours quoted in kPa.}
\label{addeffects}
\end{figure}

\begin{figure}[!ht]
%\vspace{5mm}
\centering
\includegraphics[keepaspectratio, scale=0.9]{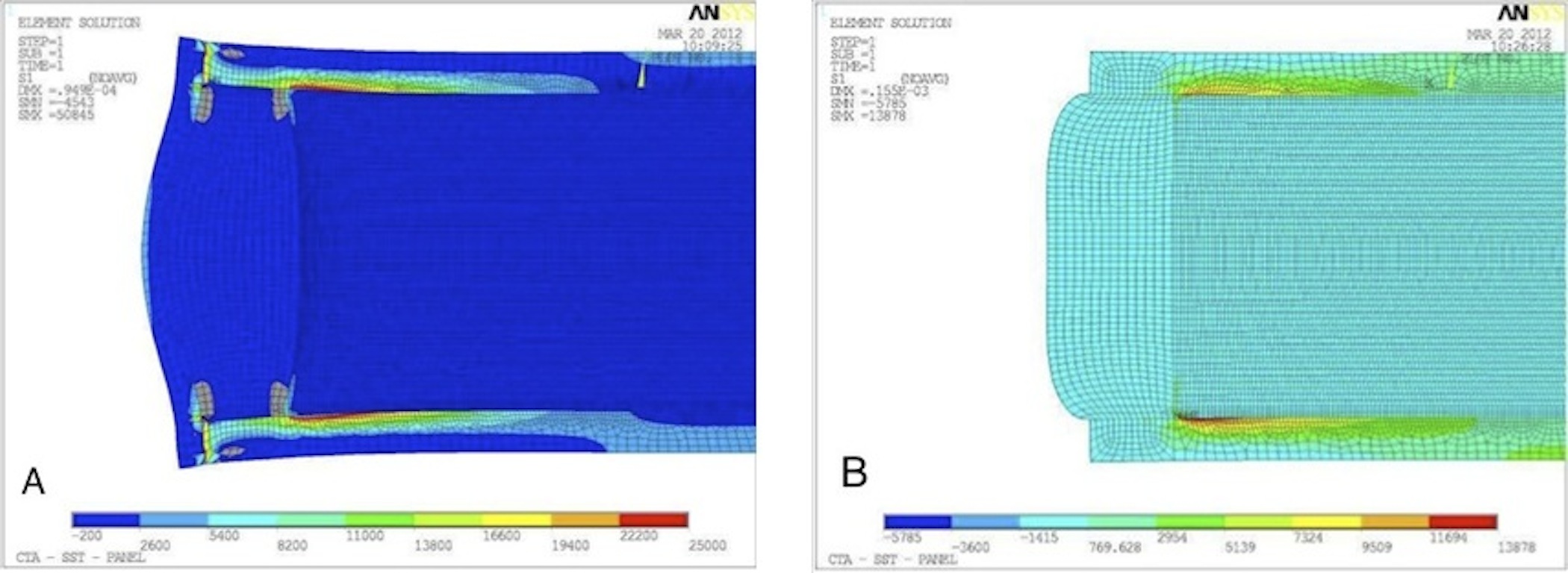}
\caption{Two examples of possible stress behavior at the edges of the mirror in case of: A- edges finished with silicone rubber sealant and a protecting PVC rim; B- edges without the PVC rim. The stress peaks differ considerably in the two implementations. Stress isocontours quoted in kPa.}
\label{edge}
\end{figure}

\noindent
Mirror size and core thickness play a substantial role in the stress peaks. The stress state into the panel structural components obviously depends on the mirror size (see \figurename~\ref{addeffects} A-B). Wind loads are also important; in particular the survival wind load represents the prominent contribution. Thus, the entity of the induced stresses poses some limits to the technology. However, in order to keep stresses within the allowable range, different solutions can be adopted. For example, the core thickness could be augmented to a certain extent, or as an alternative, a more complex support system could be required. \\
Results also show that, adopting the simplest support system consisting in three circular pucks placed at azimuthal distance of 120$^\circ$, the pucks location does not play any prominent influence on the stress peaks values, while it does in the elastic deformations that concur to deteriorate the optical prescription. The minimum values for peak-to-valley and rms errors are reported when the radial position become close to 2/3 of the panel radius (see \figurename~\ref{addeffects}~C).\\
Large temperature shifts, for example caused by daylight direct Sun heating or winter frosting, are also demanding in terms of stress loads (noting that stresses generate by thermal loads must be considered over extensive time periods.) and can also generate tensile stress of the order of few MPa. Examples in \figurename~\ref{addeffects}~D-E.\\
The mirror's edges finishing can has a major impact on the mirror because of the generation of stress caused by large thermal shifts coupled with differential CTE between the materials. Important peak values can appear for particular implementation choices, an example is shown in \figurename~\ref{edge}.\\
A more extensive description of the results can be found in~\cite{bcvstressadd}; moreover, comprehensive and detailed analyses applied to real mirrors designs are reported in~\cite{bcvmst} and~\cite{bcvsst}.\\

%Just simple preliminary considerations can be carried out.
%
%Such considerations are based on the assumptions that under the maximum loads and temperature gradients the tensile stress level in the glass have to be maintained below few MPa. LetÕs say 6-7 MPa.
%So at the moment we fixed at 3-4 MPa the maximum stress peaks induced during cold shaping. From this perspective increasing the stress admitted at CS end would means to reduce the stress share available for all other loads during operative life, making more demanding the structural panel design.

\section{The technological readiness}

\subsection{The cold-shaping implementation}
\label{impl}

\begin{figure}[!ht]
\centering
\includegraphics[keepaspectratio, scale=0.68]{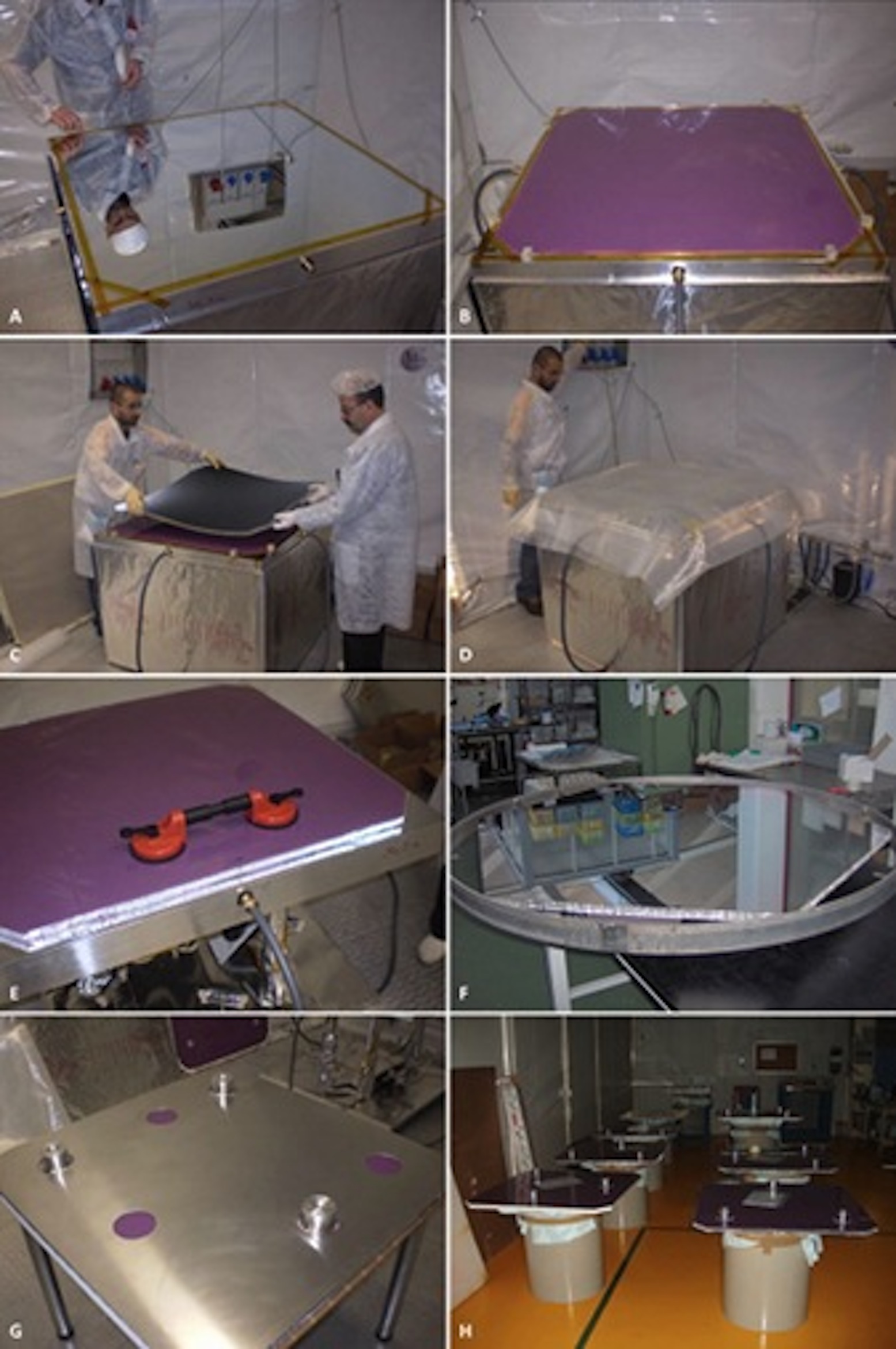}
\caption{Main steps of the manufacturing: A-B-C sandwich preparation; D- polymerization of the glue; E- sandwich release; F- coating; G-H mirror's interfaces fixation and edge finishing.}
\label{implement}
\end{figure}

The CS process has been developed by the Media Lario Technologies company under the scientific supervision of INAF-OAB~\cite{cold-mlt}.\\
The mold has to be machined once only and needs to have the same profile precision needed for the mirrors, but be the negative. The microroughness is not an issue since the CS process is not going to replicate it. The molds can be made out of metal (i.e. aluminum or steel) through (diamond) turning/milling machining or more performing processes depending on the shape accuracy requirements.\\
The remaining steps sketched in \figurename~\ref{cold} are hereafter described:
\begin{itemize}
\item a pair of glass foils and one sheet of aluminum honeycomb for the sandwich are prepared by cutting out from larger blanks. The cutting can be easily done by using shape's templates and cutters.
Glass is then carefully cleaned.
\item The first glass' foil is positioned, bent and fixed over the mold; then it is made to adhere by vacuum suction. In this way the shape of the mold is replicated; afterwards the sandwich is assembled. The connection between the honeycomb sheet and the glass foils is achieved by bonding the parts together with epoxy resin structural adhesive. Photographic images of the sandwich preparation are shown in the A-B-C panels of \figurename~\ref{implement}.
\item The resin is made polymerize with the proper curing cycle. Temperatures and timing play a role in the resulting radius of curvature and shape of the mirror, as well as the amount of glue (see \figurename~\ref{implement} D).
\item Once the polymerization has taken place, the vacuum suction is stopped and the sandwich is carefully released from the mold (see \figurename~\ref{implement} E).
\item After a deep cleaning of the front glass the reflecting coating is deposited. Methods and layers are chosen in accordance with the final application of the mirror. For example, in case of outdoor applications such as for the Cherenkov, a protective coating (e.g. Quartz) has been applied to enhance the surface strength and peak reflectivity. It also worth mentioning that, since the sandwich is kept together by the glue, attention must be paid to the heating caused by the coating process (\figurename~\ref{implement} F).
\item Finally the interfaces with the telescope supporting structure are fixed and the edges of the mirror can be sealed. This solution also ensures higher rigidity and mechanical protection of the mirror edges/corners (see \figurename~\ref{implement} G-E).
\end{itemize}

\noindent
A few remarks are here reported. Both the glass foils and the honeycomb sheets can be either off-the-shelf commercially available products or customized ones depending on the target cost and requirements of the mirrors under manufacturing. For examples, considering the relaxed optical requirements for the Cherenkov telescopes application the market provides a wide variety of technical glass having the required thickness and surface microroughness. Avoiding to use optical glass it is possible to keep the costs very low.\\
Concerning the CS process, it is clear that as the glass foils and honeycomb are pressed against the mold an elastic deformation occurs and so, at mold removal, some sort of spring back is expected. The spring back phenomenon, whose amplitude can be preventively estimated by proper analyses, should be taken into account in the mold machining in order to reduce the shape error of the panel. Nevertheless, the CS turns to be very flexible in terms of radius of curvature of the mirrors. In fact, if from a single replication mold it could be thought that only one radius of curvature should be obtained, by a careful control of the spring back effect during the gluing process, it is possible to obtain a controlled spread in the radius of curvature of the mirrors. As an example, the case of the MAGIC II project~\cite{cold-oab} is shown in~\ref{mass-prod}.

\subsection{Validation tests}
\begin{figure}[!ht]
\centering
\includegraphics[keepaspectratio, scale=0.85]{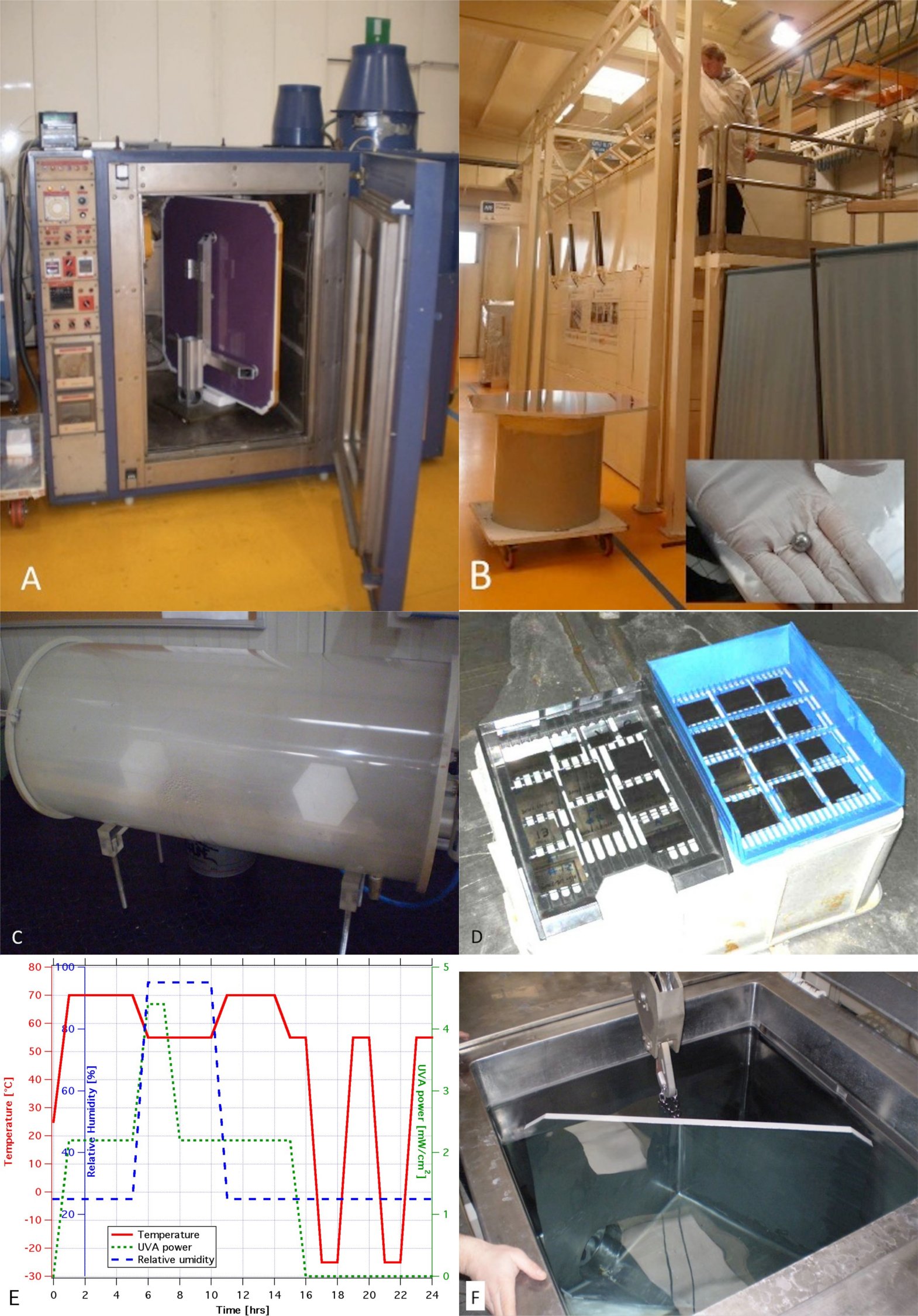}
\caption{A- Temperature cycling test; B- Mechanical impact test; C- Salt-fog test; D- Damp heat test; E- temperature, humidity and UVA power cycles for weathering tests; F- Soak test.}
\label{validation}
\end{figure}

Because of the peculiarity of IACT as briefly described in the introduction, the mirrors have been intensively tested for their robustness. The most significant results are discussed in this section.\\
One of the test performed was to investigate to what extent the performances of the mirror are affected by climatic stress (temperature cycling). The environmental durability of the coating reflective layer and of its adhesion to the glass substrate have been tested on specimens. However, such performances were assessed also on full size mirrors; furthermore the optical performance stability has been evaluated. Thermal cycles were performed ranging from -10$^\circ$C up to 60$^\circ$C. A degradation of the order of the 3\% on the focal spot dimension has been measured, while no coating removal was observed on the optical surface. \figurename~\ref{validation} A shows a mirror fitted into the climatic chamber.\\
The mirrors may occasionally be subject to mechanical impact, such as birds flying or picking on the mirrors, or possibly hail fall. The mirrors has been also tested in term of resistance to the mechanical impact stress. A steel ball has been dropped over the optical surface of the mirror accordingly with BS 7527-2.2:1991 degree of severity 1 (see  \figurename~\ref{validation} B). No evidence of damage has been observed after the test. No cracks, bumps or defects on the optical surface has been highlighted from the visual inspection performed after the impact.\\
Weathering tests, such as salt-fog, damp heat, salt-mist, humidity and UV-A irradiation, have been also performed on specimens. The tests have been lasted for 42 days with a cyclical changes of the temperature, relative humidity and UV-A irradiation power profiles as plotted in \figurename~\ref{validation} E. No measurable change in the reflectivity has been detected.
As additional test, the mirrors have been completely dipped into water for the duration of 24 hours. The weight of the mirrors before and after the test was checked: no penetration of water inside the sandwich structure has occurred.

\subsection{The mirrors production}
\label{mass-prod}
\begin{figure}[!ht]
\centering
\includegraphics[keepaspectratio, scale=1.5]{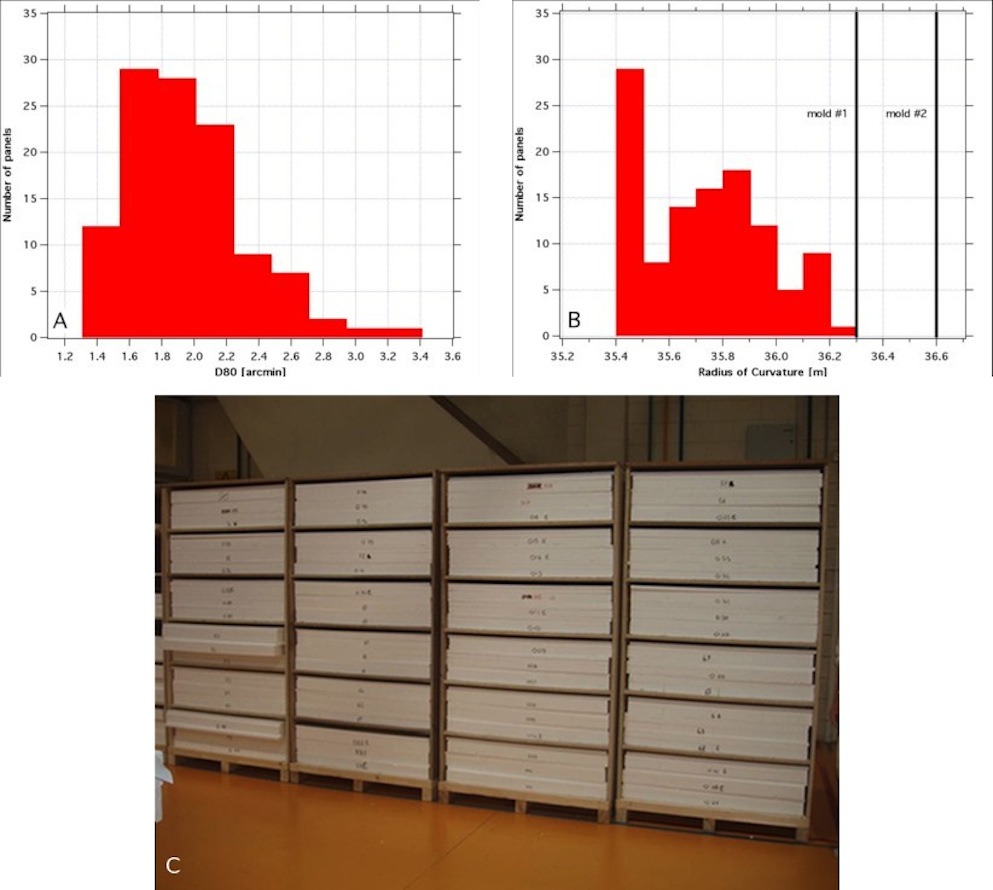}
\caption{The distributions of the D80 (A-) and the radii of curvature (B-) for the 112 mirrors produced for the MAGIC II telescope. The full production into their shipping boxes (C-).}
\label{magicstats}
\end{figure}

The possibility to exploit technologies capable of mass produce the mirrors is also a key feature for the success of the next generation of telescopes. The CS process has demonstrated to have also this potential. We report two examples of its application, both concerning the case of Cherenkov telescopes, conducted by Media Lario Technologies and INAF-OAB.\\
The MAGIC~II telescope~\cite{magici}~\cite{magicii} (see \figurename~\ref{magicghiaccio}) is the largest Cherenkov telescope in operation today as well as the largest optical light collector with imaging capabilities. It is located at the La Palma island (Canary Island) at 2200~m asl at the Observatorio del Roque de los Muchacho. MAGIC~II  has a segmented surface of nearly 240~m$^2$ composed by square tiles. The telescope has a parabolic nominal profile; it is approximated by spherical mirror segments with appropriate radius of curvature. More than 100 of those have been realized with the CS process. Each mirror has 1~m$^2$ in area and less that 10~kg in weight achieving an extremely aggressive areal density profile. The mirrors show a typical residual error of about 15~$\mu$m rms with respect to the best fitting sphere, a factor three greater with respect to the replication molds. The light concentration, defined as the circle containing the 80\% of the focused light (D80), turns to be around 2~arcmin as shown in \figurename~\ref{magicstats}~A.\\
The full production has taken less than three months and two replication molds. The plot in \figurename~\ref{magicstats}~B shows the distribution of the radii of curvature of the mirrors in comparison with the molds. The spring back effect after the vacuum suction release is well visible; however, by a careful control of the process it has been possible to produce mirrors with a dozen of different radii of curvature. The (different) radii of curvature produced were deliberately matched to the telescope requirements. This fact has permitted to minimize the number of replication molds and maximize their depreciation over the cost of each mirror. Finally, a production yield as high as the 97.4\% has been successfully achieved.

\begin{figure}[!ht]
\centering
\includegraphics[keepaspectratio, scale=1.2]{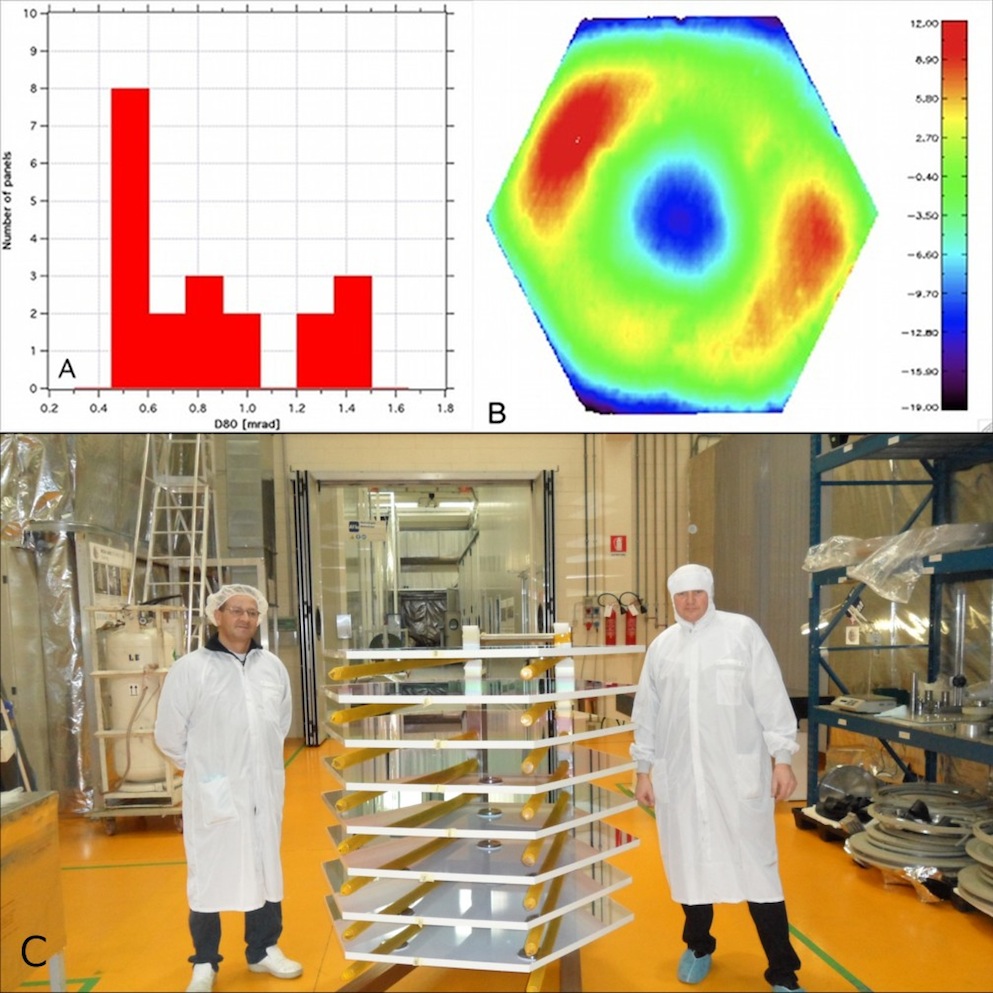}
\caption{The distributions of the D80 at the nominal radius of curvature (A-) and the typical residual errors map (B-). Part of the mirrors ready to be packed and shipped (C-).}
\label{ctamststats}
\end{figure}

\noindent
Vice versa, for the CTA-MST project~\cite{ctamlt} it has been required to demonstrate the repeatability of the CS process by delivering a set of mirrors within the specification: D80 better than 1.5~mrad at the nominal radius of curvature with a goal of 1~mrad (i.e. at the same radius of curvature for all the mirrors). A second production has been performed on a small number of prototypes. The mirrors were hexagonal in shape (1.12~m flat-to-flat) with an enlarged dimension with respect to the MAGIC~II production. As shown in \figurename~\ref{ctamststats} A, all the mirrors were compliant with the specification with fifthteen out of twenty reaching the goal. The residual errors map of the mirror surface (with respect to the best sphere) is typically better than 6~$\mu$m rms; an example is reported in \figurename~\ref{ctamststats} B.\\ 
Finally, the mass of the mirrors has been measured to be 11.75~kg and the cost of some thousands of Euro.

\section{Conclusions}
The paper presents a novel method for the realization of segmented mirror surfaces. Two thin glass foils are bent, at room temperature, over a replication mold and an honeycomb sheet is interposed between the two. A structural adhesive keeps the parts together forming a stiff and lightweight sandwich structure. This process has been called Cold Shaping.\\
The implementation is described and a number of validation tests have been discussed. The authors have also shown a couple of examples demonstrating the scalability of the process to industrial level with the production of several tens of pieces.\\
In parallel, the process has been studied also by means of finite element analyses in order to probe the wide space of the parameters. The analyses have addressed the bending limits in relation to the shape, dimension and thickness of the glass; moreover, the effects of a variety of loads (wind, temperature, gravity) have been commented.\\
Thought the validation tests performed and the application on the a real case (i.e. the MAGIC II telescope) it has been demonstrated the validity of the CS process in delivering lightweight but very robust mirrors for aggressive environments. In fact, the process has been developed for the realization of mirrors for Cherenkov telescopes for which the mechanical properties prevail on the optical ones. However, different choices on the materials, the implementation and tailored designs, could in principle lead the realization of mirrors with more demanding optical requirements.

\section*{Acknowledgment}
The authors are grateful for support from the Italian Ministry of Education, Universities, and Research (MIUR), and by the Italian National Institute for Astrophysics (INAF). The valuable collaboration with BCV progetti s.r.l. and Media Lario s.p.a is very acknowledged.

%\vspace{\baselineskip}

\section*{Biographies}

\begin{floatingfigure}{5cm}
\mbox{\includegraphics[keepaspectratio, scale=1.5]{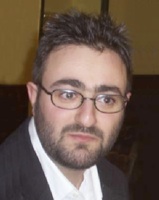}\\}
\end{floatingfigure}
\noindent
\textbf{Rodolfo Canestrari} received his masterÕs degree in astronomy in 2006 at the University of Bologna (Italy) working on multilayer coating for hard X-ray mirrors. He has continued with a PhD in astronomy and astrophysics received in 2010 at the University of Como with a thesis on thin glass sheets for innovative mirrors in astronomical applications. His main research interests are the development of mirror-manufacturing technologies and ion-beam figuring for high-precision optical components. He is heavily involved in both mirror and telescope structure development for the Cherenkov Telescope Array and ASTRI programs.\\

%\begin{figure}[!ht]
%%\vspace{5mm}
%\centering
%\includegraphics[keepaspectratio]{Canestrari.jpg}\\
%\end{figure}
%\noindent
%\textbf{Rodolfo Canestrari} received his masterÕs degree in astronomy in 2006 at the University of Bologna (Italy) working on multilayer coating for hard X-ray mirrors. He has continued with a PhD in astronomy and astrophysics received in 2010 at the University of Como with a thesis on thin glass sheets for innovative mirrors in astronomical applications. His main research interests are the development of mirror-manufacturing technologies and ion-beam figuring for high-precision optical components. He is heavily involved in both mirror and telescope structure development for the Cherenkov Telescope Array and ASTRI programs.\\

\begin{floatingfigure}{5cm}
\mbox{\includegraphics[keepaspectratio, scale=1.5]{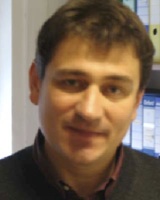}\\}
\end{floatingfigure}
\noindent
\textbf{Giovanni Pareschi} received his masterÕs degree in astronomy in 1992 at the University of Bologna (Italy), and his PhD in physics in 1996 at the University of Ferrara (Italy). From 1997 to 1998 he was an ESA postdoctoral fellow at the Danish Space Re-search Center of Copenhagen (Denmark). Since 1999 he has been an associate astronomer with INAF/Osservatorio Astronomico di Brera (Merate, Italy) and he is presently the Director of the Institute. His main field of interest is the development, implementation, and calibration of x-ray optics for astronomical missions. He is the Principal Investigator of the ASTRI program.\\

%\begin{figure}[!ht]
%%\vspace{5mm}
%\centering
%\includegraphics[keepaspectratio]{Pareschi.jpg}\\
%\end{figure}
%\noindent
%\textbf{Giovanni Pareschi} received his masterÕs degree in astronomy in 1992 at the University of Bologna (Italy), and his PhD in physics in 1996 at the University of Ferrara (Italy). From 1997 to 1998 he was an ESA postdoctoral fellow at the Danish Space Re-search Center of Copenhagen (Denmark). Since 1999 he has been an associate astronomer with INAF/Osservatorio Astronomico di Brera (Merate, Italy) and he is presently the Director of the Institute. His main field of interest is the development, implementation, and calibration of x-ray optics for astronomical missions. He is the Principal Investigator of the ASTRI program.\\

\noindent
\textbf{Giancarlo Parodi, Francesco Martelli, Nadia Missaglia and Robert Banham}: bios and photos not available\\

\end{document}